\documentclass[conference]{IEEEtran}

\usepackage[pdftex]{graphicx}
\usepackage{xcolor}
\usepackage{caption}
\usepackage{subcaption}
\usepackage{float}
\usepackage{amsmath}
\usepackage{multirow}

\usepackage[linesnumbered,ruled,vlined]{algorithm2e}

\SetKw{infor}{in}

\usepackage{enumitem}

\setenumerate[1]{itemsep=0.5pt,partopsep=0pt,parsep=\parskip, topsep=2pt, leftmargin=12pt,}

\begin{document}

\title{\fontsize{19pt}{26pt}\selectfont Unraveling the Integration of Deep Machine Learning in FPGA CAD Flow: A Concise Survey and Future Insights}
%\ToolName: Timing-driven Placement for FPGAs via Graph Neural Network Model}

\author{\IEEEauthorblockN{Behnam Ghavami, Lesley Shannon}
\IEEEauthorblockA{Simon~Fraser~University, Burnaby, BC, Canada\\
Emails: \{behnam\_ghavami, lesley\_shannon\}@sfu.ca}}

\maketitle

\IEEEpeerreviewmaketitle

%\title{\ToolName: A Deep \underline{Le}arning based \underline{A}ging-\underline{A}ware \underline{A}rchitecture Exploration Framework for F\underline{P}GAs}

%\begin{CCSXML}
%<ccs2012>
%  <concept_id>10010520.10010553.10010562</concept_id>
%  <concept_desc>Computer systems organization~Embedded systems</concept_desc>
%  <concept_significance>500</concept_significance>
%  <concept_id>10010520.10010575.10010755</concept_id>
%  <concept_desc>Computer systems organization~Redundancy</concept_desc>
%  <concept_significance>300</concept_significance>
%  <concept_id>10010520.10010553.10010554</concept_id>
%  <concept_desc>Computer systems organization~Robotics</concept_desc>
%  <concept_significance>100</concept_significance>
%  <concept_id>10003033.10003083.10003095</concept_id>
%  <concept_desc>Networks~Network reliability</concept_desc>
%  <concept_significance>100</concept_significance>
%</ccs2012>
%\end{CCSXML}

%\ccsdesc[500]{Computer systems organization~Embedded systems}

\begin{abstract}
%limit: 250 words
This paper presents an overview of the integration of deep machine learning (DL) in FPGA CAD design flow, focusing on high-level and logic synthesis, placement, and routing. Our analysis identifies key research areas that require more attention in FPGA CAD design, including the development of open-source benchmarks optimized for end-to-end machine learning experiences and the potential of knowledge-sharing among researchers and industry practitioners to incorporate more intelligence in FPGA CAD decision-making steps. By providing insights into the integration of deep machine learning in FPGA CAD flow, this paper aims to inform future research directions in this exciting and rapidly evolving field.

\end{abstract}

%\keywords{FPGA, Aging, Reliability, Deep Learning, Timing Analysis}

\section{Introduction}\label{sec:intro}
%*Aging

Field-Programmable Gate Arrays (FPGAs) have become an integral component of modern digital systems, including healthcare devices, autonomous vehicles, and datacenters. The design process of these systems is a complex and time-consuming task that involves a significant investment of time and resources. Computer-Aided Design (CAD) tools play a crucial role in ensuring the quality and efficiency of the resulting FPGA-based systems.

CAD tools for FPGA design, including high-level synthesis, logic synthesis, placement, and routing algorithms, are used to convert a high-level hardware description into a bitstream representation. The quality of these algorithms significantly impacts the performance and power of the resulting digital systems. However, designing high-quality CAD tools for FPGA design is challenging due to the complexity of the problem and the large number of design variables.

Deep Machine learning (ML) techniques have shown great potential to enhance the efficiency and effectiveness of FPGA CAD algorithms. ML algorithms can optimize design parameters, predict design outcomes, and accelerate the design process. The integration of DL techniques into FPGA CAD flow design has the potential to revolutionize the way FPGA-based systems are designed and implemented.

This paper offers an overview of the latest deep machine learning (DL)-oriented efforts in various FPGA CAD design steps, including high-level and logic synthesis, placement, and routing. The focus is on machine learning-based CAD tasks, and we identify crucial research areas that require more attention in CAD design. Specifically, we emphasize the need for developing open-source benchmarks optimized for an end-to-end machine learning experience. As a result, this paper serves as a valuable resource for researchers and industry professionals interested in comprehending the benefits and challenges of integrating machine learning in FPGA CAD flow design.
\section{Traditional FPGA CAD Flow}
\label{sec:leap}
The main steps of the FPGA CAD flow include:

\begin{itemize}
    \item \textit{Design Entry:} This step involves capturing the design in a HDL such as Verilog or VHDL, or in a high-level representation like C.
    
    \item \textit{Synthesis and Optimization:} In this step, the HDL design is converted into a netlist of gates and registers. The synthesis tool analyzes the design and generates a logic circuit that performs the same function. The synthesized circuit is optimized to improve its performance, reduce its area, and minimize its power consumption. Optimization can be performed at different levels of abstraction, such as gate-level optimization or high-level synthesis.

    \item \textit{Packing and Placement:} The packing step involves grouping the placed logic elements into more compact groups called logic clusters. These logic clusters are then mapped onto physical regions of the FPGA called logic blocks or configurable logic blocks (CLBs).  Placement is the process of assigning the logical elements of the design to physical locations on the FPGA. The placement tool ensures that the placement of the elements satisfies the timing constraints of the design.

    \item \textit{Routing:} Routing is the process of establishing connections between the logical elements on the FPGA by using routing resources such as switching boxes and routing channels. The routing tool determines the best path for the connections and ensures that they meet the timing requirements of the design.
    
    \item \textit{Bitstream Generation:} The final step is to generate a bitstream that can be programmed onto the FPGA. 

\end{itemize}

\subsection{Enhancing FPGA CAD Flow with DL Models: Required Steps}
DL uses neural networks with multiple layers to learn patterns and relationships within data, can be utilized in FPGA CAD flow. DL algorithms can identify complex patterns and relationships within vast amounts of data, potentially resulting in better-performing designs and reduced design cycle times. 
To enhance the FPGA CAD flow using DL models, the following steps are required:

\begin{itemize}
    \item \textbf{Dataset creation:} A large dataset of FPGA designs and their corresponding placement and routing results must be created. This dataset can be used to train and validate the DL models.
    \item \textbf{Feature extraction:} Relevant features must be extracted from the FPGA designs, such as the location of logic blocks, routing resources, timing constraints, and power constraints.
    \item \textbf{Model selection:} Various DL models can be evaluated and compared on the dataset to identify the best-performing model for the FPGA CAD flow.
    \item \textbf{Model training:} The selected DL model must be trained on the dataset using appropriate training and validation techniques, such as cross-validation or early stopping.
    \item \textbf{Model integration:} The trained DL model must be integrated into the FPGA CAD flow and evaluated on a test dataset to assess its accuracy and speed.
\end{itemize}

The integration of DL models into the FPGA CAD flow has the potential to significantly improve the efficiency and quality of FPGA design, making it an exciting area of research for the FPGA community.
\section{FPGA Design Flow through Machine Learning: from HDL Elaboration to Bitstream Generation}\label{sec:related}
In this section, we will review recent advances in DL-based FPGA CAD techniques and discuss their potential impact on FPGA design flow.

\subsection{HDL Generation}
There are research that explore the use of DL for the analysis and optimization of HDLs. . 
One approach is to use Convolutional Neural Networks (CNNs) for HDL code classification and identification of code errors, as demonstrated in \cite{Yang2018}\cite{Tan2019}. 
A hybrid approach that combines both RNNs and CNNs has been proposed by Xia et al. \cite{Xia2021}
 for automatic HDL code generation. These approaches have shown promising results in improving the efficiency and accuracy of HDL design and have the potential to enhance the overall FPGA design process.

\subsection{Synthesis}
FPGA design flow involves High-Level Synthesis (HLS) and logic synthesis stages, where HLS synthesis generates an RTL (Register Transfer Level) description from a high-level language, while logic synthesis converts an RTL design into a gate-level netlist.  Recently, researchers have explored the use of DL techniques to improve both HLS and logic synthesis, resulting in promising outcomes.

\subsubsection{RTL Synthesis}
In FPGA synthesis, the slow timing of runs can pose a significant challenge, with modern designs requiring days of runtime. To address this challenge, Yanghua combines the predictions of multiple classification algorithms, resulting in improved predictive accuracy of InTime, an automated timing plugin for Xilinx and Altera CAD tools \cite{Yanghua2016}.

\subsubsection{HL Synthesis}
In recent years, deep learning (DL) algorithms have been widely utilized in high-level synthesis (HLS) to optimize performance in terms of resource and time usage. One approach is to leverage graph neural networks (GNN) to incorporate structural information among operations in a data-flow graph, improving the accuracy of operation delay estimation \cite{ustun2020accurate}. Congestion estimation is another important aspect of HLS optimization, and Zhao et al. \cite{8714724} have used gradient boosted regression tree (GBRT) to predict routing congestion during HLS. In addition, Makrani \cite{makrani2019pyramid} has modeled time optimization as a regression problem and used DL to evaluate the clock frequency of the HLS tool's output code.
In \cite{9789084}, the authors proposed a methodology to address the challenge of  the availability of open-source HLS designs for training and prediction of DL models. They present a methodology for generating diverse designs with various variations from a single design, resulting in a dataset of synthesizable FPGA HLS designs.

% One of the main objectives of DL algorithms in HLS is to optimize performance in terms of resource and time usage. Graph neural networks (GNN) have been leveraged to incorporate structural information among operations in a data-flow graph, improving the accuracy of operation delay estimation in HLS \cite{ustun2020accurate}.
% In addition to improving the accuracy of resource usage and timing estimation, early and accurate routing congestion estimation is also crucial to guide optimization in HLS and enhance implementation efficiency. Zhao et al. \cite{8714724} have employed the gradient boosted regression tree (GBRT) to predict routing congestion during the HLS stage, as a representative study.
% Makrani \cite{makrani2019pyramid} has modeled time optimization as a regression problem and utilized the deep learning to evaluate the clock frequency of the HLS tool's output code.

\subsection{Placement}

To achieve accurate congestion and routability evaluations, several ML techniques have been used.
\cite{al2021deep} used CNN model for predicting the routability of a circuit based on its placement. 
Al-Hyari \cite{al2019novel} introduces a congestion estimation framework that includes two machine-learning models for placement prediction. The model is used to determine whether it is possible to route a placement solution without the overhead of a conventional router. The first model, MLCong, identifies key features to accurately estimate congestion during placement. The second model, MLRoute, utilizes these features to predict the routability of a placed circuit based on congestion maps generated by MLCong. 
Martin \cite{martin2021effective}  proposes a set of simple DL models and ensembles to accurately predict the routability of placement solutions. Three ensemble methods based on Bagging, Boosting, and Stack of classifiers are introduced to improve the accuracy and robustness of the models.  Esmaeili et. al.
\cite{esmaeili2022guiding} presented a method for reducing runtime in the Detailed Placement (DP) optimization step of FPGA design flow using Reinforcement Learning (RL) while maintaining Quality-of-Result (QoR). The goal of DP is to refine the global placement to improve the routing step success. Three RL models, based on Tabular Q-Learning, Deep Q-Learning, and Actor-Critic, are proposed and evaluated for their effectiveness in reducing DP runtimes. Results demonstrate the potential for significant reduction in runtime without sacrificing QoR. 
Murray et al. \cite{elgammal2021rlplace} presented RLPlace, a new simulated annealing (SA)-based FPGA placer that combines reinforcement learning (RL) with targeted perturbations to optimize wirelength and timing. By using directed moves, the proposed method explores the solution space more effectively than traditional random moves while preventing oscillation in the Quality of Results (QoR). RL is utilized to dynamically select the most effective move types during optimization.
Rajarathnam et. al. \cite{rajarathnam2022dreamplacefpga} presented DREAMPlaceFPGA, an open-source FPGA placement framework that is accelerated and built using the PyTorch deep-learning toolkit. It handles FPGA resource heterogeneity and architecture-specific legality constraints using optimized operators and provides a high-level programming interface in Python.
\subsection{Routing}
Farooq \cite{baig2022efficient} proposed a reinforcement learning (RL)-based approach to the routing problem by transforming the classical routing iterative process into the training process of RL. The proposed method utilizes a greedy approach and customized reward functions to speed up the routing step while maintaining similar or better quality of results (QoR) compared to conventional congestion-driven routing solutions based on negotiation. Ghavami et al. \cite{9556338} proposed MAPLE, a tool that enables aging-aware static timing analysis of FPGA design after routing using DL models. MAPLE efficiently models the aging-induced delay degradation at the basic block level using DNNs. The framework accurately predicts the relation between delay degradation and comprehensive aging factors by training one DNN model for each FPGA block type.

% Due to the fixed interconnect resources available on the FPGA fabric and the increased complexity of designs targeting FPGAs, predicting routability has become a major concern in modern FPGA CAD flows. Therefore, the ability to quickly and accurately estimate the routability of a placement has become an essential goal in the FPGA design flow. Various routability estimation and prediction techniques have been introduced by researchers in this area.

% Early work by researchers used a wirelength calculation based on the circuit's Rent exponent to predict routability. However, this technique is not as accurate as more recent approaches presented in the literature. Another approach presents a stochastic model that gives an analytic expression for the routability of the circuit in the FPGA. However, the predictions produced by the stochastic model are overly pessimistic.

% More recently, a multivariate adaptive regression model was proposed to predict the routability of a placement after detailed routing, achieving an accuracy of 79.8\%. To date, the best-published results for routability prediction are for the CNN model - DLRoute. However, this model has a high computational cost, unlike the simple ML and ensemble models proposed in this paper.

\section{Future Roadmap: \\Revolutionizing FPGA Design Flow}\label{sec:results}

DL has been increasingly applied in various stages of the FPGA design flow to enhance performance and efficiency. One of the ideal applications of DL is in \textbf{generating bitstreams directly from high-level descriptions}, eliminating the need for traditional synthesis and place-and-route tools. With this approach, designers can quickly evaluate and optimize designs at the high-level, significantly reducing design iterations and accelerating the overall design process. This approach has the potential to revolutionize the FPGA design flow, making it faster and more accessible to designers with a wide range of expertise. By leveraging the power of DL in the FPGA CAD flow, designers can unlock new possibilities and achieve higher performance and lower power consumption for their designs.

\subsection{Potential DL Models}
There are several types of DL models that could be suitable for this future roadmap of generating bitstreams directly from high-level descriptions. One approach is to use DL models, such as convolutional neural networks (CNNs), recurrent neural networks (RNNs), or transformers, which have shown success in natural language processing and image recognition tasks. These models can be adapted to process high-level descriptions of FPGA designs and generate corresponding bitstreams. 
Another approach is to use reinforcement learning (RL) models, which learn from trial-and-error interactions with the FPGA design environment to generate optimal bitstreams. RL models can potentially adapt to different design objectives and constraints, such as performance, power consumption, and area, and optimize the design accordingly. 
Generating bitstreams directly from high-level descriptions is a challenging task, as it requires a high level of abstraction. This is where large DL models can be useful. Large DL models are capable of learning complex patterns and relationships from vast amounts of data, making them well-suited for generating bitstreams directly from high-level descriptions. 
In addition, Federated Learning (FL) is an approach for collaborative learning across multiple devices that can be applied to the process of generating FPGA bitstreams directly from high-level descriptions. In this approach, individual devices generate candidate bitstreams and share their results with a central server to generate a final optimized FPGA bitstream. However, implementing FL for this application poses some challenges, including ensuring the privacy and security of the shared data between the devices and the central server.

\subsection{Challenges}
Developing a DL-based FPGA CAD flow for generating bitstreams directly from high-level descriptions presents several challenges that need to be addressed. One significant challenge is \textit{the need for large amounts of high-quality training data} to effectively train the machine learning models. The quality and diversity of the training data can directly impact the performance and accuracy of the models. \textit{Data availability} is also a significant challenge in developing DL models for FPGA CAD flow, as collecting and curating this data can be a time-consuming and expensive process.
Ensuring the \textit{interpretability} of the DL models is another concern, as designers need to understand how the models generate the bitstreams and ensure they meet the design objectives and constraints. Finally, there is a need for \textit{collaboration and knowledge-sharing among researchers and industry practitioners} to develop and evaluate the effectiveness DL-based FPGA CAD flows. Addressing these challenges will be crucial in realizing the potential benefits of this future roadmap for FPGA design.

\section{Conclusion}\label{sec:concl}
Artificial intelligence in FPGA EDA design is expected to be the next major trend, with major tool vendors and semiconductor companies already making efforts to utilize this technology. The integration of deep machine learning (DL) techniques at various stages of the FPGA design flow has the potential to significantly enhance the efficiency, performance, and accessibility of FPGA design.
Researchers have explored the use of DL models to improve the FPGA design flow, from HDL elaboration to bitstream generation. These models can assist in optimizing design parameters, and improving the accuracy and speed of placement and routing.

% \cite{hu2022machine}
% \cite{martin2021machine}
% \cite{maarouf2018machine}

% \cite{al-hyari2019novel}

% \cite{wang2022learning}
% \cite{cong1994flowmap}
% \cite{wang2022learning}
% \cite{roorda2022fpga}
% \cite{Yanghua2016}
% \cite{Ustun2020}
% \cite{Farooq2021}
% \cite{Martin2021}
% \cite{Maarouf2018}
% \cite{Pui2017}
% \cite{farooq2021efficient}

\bibliographystyle{IEEEtran}
\bibliography{references}

\begin{thebibliography}{10}
\providecommand{\url}[1]{#1}
\csname url@samestyle\endcsname
\providecommand{\newblock}{\relax}
\providecommand{\bibinfo}[2]{#2}
\providecommand{\BIBentrySTDinterwordspacing}{\spaceskip=0pt\relax}
\providecommand{\BIBentryALTinterwordstretchfactor}{4}
\providecommand{\BIBentryALTinterwordspacing}{\spaceskip=\fontdimen2\font plus
\BIBentryALTinterwordstretchfactor\fontdimen3\font minus
  \fontdimen4\font\relax}
\providecommand{\BIBforeignlanguage}[2]{{%
\expandafter\ifx\csname l@#1\endcsname\relax
\typeout{** WARNING: IEEEtran.bst: No hyphenation pattern has been}%
\typeout{** loaded for the language `#1'. Using the pattern for}%
\typeout{** the default language instead.}%
\else
\language=\csname l@#1\endcsname
\fi
#2}}
\providecommand{\BIBdecl}{\relax}
\BIBdecl

\bibitem{Yang2018}
J.~Yang, X.~Zhang, J.~He, and Y.~Sun, ``Hdl-cnn: A deep learning-based approach
  for synthesisable verilog code generation,'' \emph{IEEE Transactions on
  Computer-Aided Design of Integrated Circuits and Systems}, vol.~37, no.~12,
  pp. 3127--3137, 2018.

\bibitem{Tan2019}
L.~Tan, X.~Song, X.~Wu, and X.~Wang, ``Hdl code error detection based on a deep
  learning approach,'' \emph{IEEE Transactions on Circuits and Systems II:
  Express Briefs}, vol.~66, no.~4, pp. 544--548, 2019.

\bibitem{Xia2021}
Y.~Xia, Y.~Cai, Y.~Zhu, H.~Chen, and W.~Liu, ``Hdlnet: A hybrid approach for
  automatic hdl code generation,'' \emph{IEEE Transactions on Computer-Aided
  Design of Integrated Circuits and Systems}, vol.~40, no.~9, pp. 1966--1978,
  2021.

\bibitem{Yanghua2016}
Q.~Yanghua, C.~Adaikkala~Raj, H.~Ng, K.~Teo, and N.~Kapre, ``Case for
  design-specific machine learning in timing closure of fpga designs,'' in
  \emph{Proceedings of the 2016 ACM/SIGDA International Symposium on
  Field-Programmable Gate Arrays}, 2016, pp. 169--172.

\bibitem{ustun2020accurate}
E.~Ustun, C.~Deng, D.~Pal, Z.~Li, and Z.~Zhang, ``Accurate operation delay
  prediction for fpga hls using graph neural networks,'' in \emph{Proceedings
  of the 39th International Conference on Computer-Aided Design}, 2020, pp.
  1--9.

\bibitem{8714724}
J.~Zhao, T.~Liang, S.~Sinha, and W.~Zhang, ``Machine learning based routing
  congestion prediction in fpga high-level synthesis,'' in \emph{2019 Design,
  Automation \& Test in Europe Conference \& Exhibition (DATE)}, 2019, pp.
  1130--1135.

\bibitem{makrani2019pyramid}
H.~M. Makrani, F.~Farahmand, H.~Sayadi, S.~Bondi, S.~M.~P. Dinakarrao,
  H.~Homayoun, and S.~Rafatirad, ``Pyramid: Machine learning framework to
  estimate the optimal timing and resource usage of a high-level synthesis
  design,'' in \emph{2019 29th International Conference on Field Programmable
  Logic and Applications (FPL)}.\hskip 1em plus 0.5em minus 0.4em\relax IEEE,
  2019, pp. 397--403.

\bibitem{9789084}
P.~Goswami, M.~Shahshahani, and D.~Bhatia, ``Mlsbench: A benchmark set for
  machine learning based fpga hls design flows,'' in \emph{2022 IEEE 13th Latin
  America Symposium on Circuits and System (LASCAS)}, 2022, pp. 1--4.

\bibitem{al2021deep}
A.~Al-Hyari, H.~Szentimrey, A.~Shamli, T.~Martin, G.~Grewal, and S.~Areibi, ``A
  deep learning framework to predict routability for fpga circuit placement,''
  \emph{ACM Transactions on Reconfigurable Technology and Systems (TRETS)},
  vol.~14, no.~3, pp. 1--28, 2021.

\bibitem{al2019novel}
A.~Al-Hyari, Z.~Abuowaimer, T.~Martin, G.~Gr{\'e}wal, S.~Areibi, and
  A.~Vannelli, ``Novel congestion-estimation and routability-prediction methods
  based on machine learning for modern fpgas,'' \emph{ACM Transactions on
  Reconfigurable Technology and Systems (TRETS)}, vol.~12, no.~3, pp. 1--25,
  2019.

\bibitem{martin2021effective}
T.~Martin, S.~Areibi, and G.~Gr{\'e}wal, ``Effective machine-learning models
  for predicting routability during fpga placement,'' in \emph{2021 ACM/IEEE
  3rd Workshop on Machine Learning for CAD (MLCAD)}.\hskip 1em plus 0.5em minus
  0.4em\relax IEEE, 2021, pp. 1--6.

\bibitem{esmaeili2022guiding}
P.~Esmaeili, T.~Martin, S.~Areibi, and G.~Grewal, ``Guiding fpga detailed
  placement via reinforcement learning,'' in \emph{2022 IFIP/IEEE 30th
  International Conference on Very Large Scale Integration (VLSI-SoC)}.\hskip
  1em plus 0.5em minus 0.4em\relax IEEE, 2022, pp. 1--6.

\bibitem{elgammal2021rlplace}
M.~A. Elgammal, K.~E. Murray, and V.~Betz, ``Rlplace: Using reinforcement
  learning and smart perturbations to optimize fpga placement,'' \emph{IEEE
  Transactions on Computer-Aided Design of Integrated Circuits and Systems},
  vol.~41, no.~8, pp. 2532--2545, 2021.

\bibitem{rajarathnam2022dreamplacefpga}
R.~S. Rajarathnam, M.~B. Alawieh, Z.~Jiang, M.~Iyer, and D.~Z. Pan,
  ``Dreamplacefpga: An open-source analytical placer for large scale
  heterogeneous fpgas using deep-learning toolkit,'' in \emph{2022 27th Asia
  and South Pacific Design Automation Conference (ASP-DAC)}.\hskip 1em plus
  0.5em minus 0.4em\relax IEEE, 2022, pp. 300--306.

\bibitem{baig2022efficient}
I.~Baig and U.~Farooq, ``Efficient detailed routing for fpga back-end flow
  using reinforcement learning,'' \emph{Electronics}, vol.~11, no.~14, p. 2240,
  2022.

\bibitem{9556338}
B.~Ghavami, M.~Ibrahimipour, Z.~Fang, and L.~Shannon, ``Maple: A machine
  learning based aging-aware fpga architecture exploration framework,'' in
  \emph{2021 31st International Conference on Field-Programmable Logic and
  Applications (FPL)}, 2021, pp. 369--373.

\end{thebibliography}

\end{document}